# Understanding the Nature of Vibro-Polaritonic States in Water and Heavy Water

*Akhila Kadyan, Monu P. Suresh, Ben Johns and Jino George\**

*Department of Chemical Sciences, Indian Institute of Science Education and Research (IISER) Mohali, Punjab-140306, India.*
\*Email: jgeorge@iisermohali.ac.in

**ABSTRACT:**

One of the most popular subjects now being researched in molecular science is strong light-matter coupling. The introduction of vibrational strong coupling and the formation vibro-polaritonic states tend to modify chemical reactivity, energy transfer, and many other physical properties of the coupled system. This is achieved without external stimuli and is very sensitive to the vibrational envelope of the molecular transition. Water is an excellent vibrational oscillator, which is being used for similar experiments. However, the inhomogeneously broad OH/OD stretching vibrational band make it complicated to characterize the vibro-polaritonic states spectroscopically. In this paper, we performed vibrational strong coupling and mapped the evolution of vibro-polaritonic branches from low to high concentration of $H_2O$ and measured the on-set of strong coupling. The refractive index dispersion is correlated with the cavity tuning experiments. These results are further compared with transfer matrix simulations. Simulated data deviate as noted in the dispersion spectra as the system enters into ultra-strong coupling condition. A simple oscillator strength correction is made to include the self-dipolar interaction. Hopfield coefficients are also calculated, showing that even at ±400 cm$^{-1}$ detuning, the vibro-polaritonic states still possess light and matter components. Here, we systematically varied the concentration of $H_2O$ and mapped the weak, intermediate, and strong coupling regimes to understand the role of inhomogeneously broad OH/OD stretching vibrational band. Our finding may help to better understand the role of $H_2O$/$D_2O$ strong coupling in the recent polaritonic chemistry experiments.

Strong light-matter coupling is a distinct way of interaction between light and matter that differs from the conventional absorption and emission processes. Strong coupling reshuffles the actual energy of the parent molecule through the formation of hybrid light-matter states. This can be achieved by the interaction of a molecular transition with the confined electromagnetic field of a Fabry-Perot (FP) cavity. When the rate of energy exchange between the confined field and molecular state exceeds all other dissipation processes, the system enters into strong coupling, giving rise to the formation of two new hybrid light-matter states called polaritonic states.[1] The energy separation between these newly formed polaritonic states is termed Rabi splitting energy. The most intriguing aspect of quantum electrodynamics is that excitations are possible in the dark due to the presence of vacuum fluctuations (zero-point energy) of the electromagnetic field, resulting in vacuum Rabi splitting. The first experimental observation of strong coupling is demonstrated by Haroche and co-workers, where they coupled a Rydberg atom with a cavity photon.[2] Following Haroche experiments, numerous molecular and material research works were performed by coupling the electronic transition of a molecule with the cavity photon. Formation of polaritonic states results in the modification of several properties associated with molecules and materials, such as energy transfer rates,[3],[4],[5], charge transport,[6],[7],[8],[9],[10], radiative and non-radiative decay rates of the excited state, [11],[12],[13],[14], water splitting,[15] and chemical reactions rate.[16],[17],[18] The strong coupling regime is not limited to electronic states of molecules but can be expanded to vibrational states opening a realm of possibilities.

Water is among the most ubiquitous substances that control several chemical and biological processes; it takes the role of solvent, reactant, and transport medium in many reactions. It also shows many anomalous properties like maximum density at 4°C and decrease in density upon solidification, high specific heat capacity, high surface tension, high boiling point, high heat of vaporization, etc., which makes water even more fascinating to researchers.[19] The key cause for such anomalies exists in the structure of water consisting of hydrogen bonding interactions. In liquid water, the hydrogen bonds between water molecules are constantly forming and breaking. This dynamic behavior leads to a highly distorted long-range arrangement of water molecules. However, on a local scale, a tetrahedral arrangement is observed in which four neighboring water molecules are connected through hydrogen bonding network.[20] Extended hydrogen bonding network makes the OH stretching band inhomogeneously broad; hence, clean spectroscopic analysis is a herculean task. Vibrational strong coupling (VSC) can be achieved with OH stretching band; however, its broad envelope is debated among the strong coupling community. Here, we try to understand the nature of vibro-polaritonic states formed from water and heavy water.

There are several attempts to produce vibro-polaritonic states from solid and liquid phases of organic and inorganic molecules.[21],[22],[23],[24],[25] Vibrational strong coupling (VSC) has also been shown to cause changes in the physical and chemical properties of molecules. It has been demonstrated experimentally that VSC can accelerate or retard a chemical process.[26],[27] The resonance nature of polaritonic chemistry is tested in few theoretical models and the modification of internal vibrational relaxation pathways are also proposed.[28],[29],[30],[31] Apart from chemical reactivity, several other properties are demonstrated to get modified under VSC, like ionic conductivity,[32] crystallization process,[33] self-interaction of vibrational transition dipoles,[34] ferromagnetism,[35] solvent polarity,[36] and self-assembly of molecules.[37] Large molecules like enzymes can undergo cooperative VSC and are shown to modify the enzyme activity.[38],[39],[40],[41] In most of these reactions, VSC of water is employed as it acts as a reactant as well as a solvent. However, only a few experimental attempts are reported to understand the nature of vibro-polaritonic states in water.[42],[43] These experiments show that VSC of water results in large Rabi splitting compared to many other molecular systems. But still, a clear picture of such large Rabi splitting and the nature of vibro-polaritonic states of water is not well understood. In this work, we perform VSC of both water and heavy water by coupling the OH and OD stretching bands with an infrared FP cavity. We studied the dispersion of polaritonic states and compared the Hopfield coefficients associated with them to the index dispersion of water in the relevant energy range across OH/OD stretching band in order to understand the origin of extremely high Rabi splitting energy. Further, we investigated the concentration dependence, collective nature of strong coupling, and summarized it in terms of various coupling regimes.

As mentioned previously, inhomogeneously broad OH and OD stretching bands are recorded using a demountable microfluidic cell (figure 1). The cell consists of two $BaF_2$ substrates without spacer that gives nearly 1 to 2 $\mu$m physical separation. $H_2O$ and $D_2O$ show very strong OH/OD stretching bands at 3408 cm$^{-1}$ and 2507 cm$^{-1}$, respectively. Spectral transmittance obtained these cells are further used for TMM fitting. All the information related to multi-Lorentzian fitting of the absorption envelope is given in the SI (section 2). The smaller pathlength of the cell allow us to measure the experimental transmittance of the OH/OD vibrations and are found to be very broad, as shown in figure 1b. The inhomogeneity in water stretching band arises for many reasons. This is primarily due to hydrogen bonding and the mixing of symmetric stretching, asymmetric stretching, and the first overtone of bending modes. OH/OD stretching bands are very strong compared to bending bands, and are used for the strong coupling experiments. All the experimental cavity studies are conducted using mylar spacer (6 μm). The asymmetric envelope of OH stretching mode results in the formation of vibro-polaritonic states with differences in their intensity. Here, upper vibro-polaritonic state (VP+) comes out to be of higher intensity than the lower vibro-polaritonic state (VP-). The system is in strong coupling regime, and the full-width half maximum (FWHM) of the OH stretching band and the 7$^{th}$ cavity mode are much smaller (463 and 135 cm$^{-1}$, respectively) compared to the Rabi splitting energy (740 cm$^{-1}$). The generally employed criterion for strong coupling is that its Rabi splitting energy must be greater than the average of the FWHM of the bare cavity mode and the bare vibrational band. Figure 1(c) shows the VSC of the OD stretching band of pure D$_2$O with the 5$^{th}$ mode of the cavity having 7.5 μm pathlength. Such thick cavity can support many modes; the energy (in cm$^{-1}$) of any mode is given by relation $\bar{\upsilon} = \frac{m 10^4}{2nl}$, where $m$ is the mode order, $n$ is the refractive index of the medium, $l$ is the cavity pathlength (in μm). VSC of OD stretching band resulting in the formation of the VP+ at 2776 cm$^{-1}$ and VP- at 2247 cm$^{-1}$, giving the Rabi splitting energy of 529 cm$^{-1}$. In case of D$_2$O, the OD stretching band has FWHM of 338 cm$^{-1}$, and the cavity mode FWHM is ~131 cm$^{-1}$ giving an average FWHM of 235 cm$^{-1}$. Thus, the OD stretching band of pure D$_2$O is clearly in the strong coupling regime. In the case of both H$_2$O and D$_2$O, the stretching band is an asymmetric line shape, due to which the vibro-polaritonic bands have a huge difference in their intensity, and this is particularly true in the case of H$_2$O as the VP- state is still at the broad absorption tail of the OH stretching band.

To get a clear picture of the nature of vibro-polaritonic states, VSC of a diluted solution of H$_2$O in D$_2$O is done. 9% H$_2$O solution (v/v%) is used for coupling OH stretching band with the cavity photon (figure 2). Cavity tuning experiments are carried out where the energy of the cavity mode coupled to the OH stretching band changed between the ON and OFF-resonance conditions. ON-resonance condition means the energy of the cavity mode exactly coincides with the vibrational energy of the OH stretching band. When the energy of the cavity mode does not match exactly with the stretching band but is shifted towards higher or lower energies is referred to as the OFF-resonance condition. Tuning experiments are performed by physically changing the separation between the two-gold coated BaF$_2$ substrates by manually adjusting the screws of the demountable microfluidic cell assembly. Since the Mylar spacer used to create the separation between the substrates is elastic in nature, it enables us to change the pathlength. The experimental dispersion of VP+ and VP- formed by VSC of 9% $H_2O$ solution is shown in figure 2(a). Due to experimental limitation, the dispersion measurement is limited to only a small change in the pathlength; transfer matrix method (TMM) simulations are performed to get the broad range dispersion of the states as shown as 3D color plot in figure 2(b). TMM simulation gives information about the Rabi splitting of coupling 6-8 modes consecutively of the cavity. The Rabi splitting of the system is intact while moving from one cavity mode to the other (figure S5 and Table S3). The fitting parameter is used to obtain the cavity dispersion plot, which matches the experimental result. A clear anti-crossing behavior for vibro-polaritonic states is observed, indicating that the system is in the strong coupling regime.

The Kramers-Kronig relation, which connects the real and imaginary parts of an absorption envelope, is used here to determine the refractive index dispersion around the molecular absorption region.[44] For $H_2O$ and $D_2O$, the real part of the refractive index is extracted as a function of energy (or wavenumbers in $cm^{-1}$), as shown in figures 3(a), and 3(c). The index dispersion around the absorption band maximum of the stretching band extends to a broad range of wavenumbers in both $H_2O$ and $D_2O$. Further, the cavity dispersion experiments are performed by varying the pathlength. The experimental dispersion of VP+ and VP- as well as the uncoupled higher and lower energy modes for $H_2O$ and $D_2O$, are represented as magenta circles in figures 3(b), and 3(d), respectively. Transfer matrix method (TMM) simulations are also performed to get the broad range dispersion as the experimental tuning of pathlength is limited to 1 $\mu$m. When it comes to pure $H_2O$ and $D_2O$, the TMM exhibits a smaller splitting than the experiment; hence an oscillator strength correction is introduced in the TMM calculations to match it with experimentally observed Rabi splitting energy (SI; section 2). However, in the prior situation of 9% $H_2O$, the TMM fitting matches the Rabi splitting energy obtained in the experiments.

To further understand the nature of vibro-polaritonic states, Hopfield coefficients are calculated. Hopfield coefficients represent the fraction of matter and photon contents in the polaritonic states at different detuning positions and can be calculated by constructing a $2X2$ matrix as follows:

$$\begin{bmatrix} E_C & \hbar\Omega_R/2 \\ \hbar\Omega_R/2 & E_X \end{bmatrix} \begin{bmatrix} \alpha \\ \beta \end{bmatrix} = E_\pm \begin{bmatrix} \alpha \\ \beta \end{bmatrix}$$

where, $E_C$ is cavity mode energy, $E_X$ is molecule transition energy (here, OH/OD stretching), $\hbar\Omega_R$ is Rabi splitting energy and $E_+$ and $E_-$ corresponding to upper and lower vibro-polaritonic states, respectively.[45] Diagonalization of the matrix gives the energy of polaritonic states as:

$$E_\pm = \frac{E_C + E_X}{2} \pm \sqrt{(\hbar\Omega_R)^2 + (E_C - E_X)^2}$$

And Hopfield coefficients $|\alpha|^2$ and $|\beta|^2$ corresponding to photon and matter fractions, respectively, comes out as:

$$|\alpha|^2 = \frac{1}{2}\left[1 + \frac{\Delta E}{\sqrt{\Delta E^2 + (\hbar\Omega_R)^2}}\right] ; |\beta|^2 = \frac{1}{2}\left[1 - \frac{\Delta E}{\sqrt{\Delta E^2 + (\hbar\Omega_R)^2}}\right]$$

where, $\Delta E = E_C - E_X$ is the detuning energy.

The calculated Hopfield coefficients for the lower polaritonic states of $H_2O$ and $D_2O$, respectively, are shown in figures 4(a) and 4(b). These are estimated by taking a smaller cavity ~1 $\mu$m pathlength in TMM, avoiding the involvement of higher and lower cavity modes. Please note that multimode mixing is possible if one use thicker cavities for estimating the mixing fraction in a large detuning range. Whereas, the experimental data points are collected for 7.5 $\mu$m cavity (7th mode). Additionally, both positive and negative detuning data are acquired with $\Delta E$ range of ±400 cm$^{-1}$ using TMM calculation. At the ON-resonance condition, both photon and matter fractions are 0.5. The polaritonic state does not decouple from the cavity photon even at a detuning of ±400 cm$^{-1}$. In the case of $H_2O$, for a detuning of ±400 cm$^{-1}$, VP- still has 0.7 matter fraction and 0.3 photon fraction. For $D_2O$, $\Delta E$ at ±400 cm$^{-1}$, VP- has 0.75 matter fraction and 0.25 photon fraction. This indicates that the stretching band of both $H_2O$ and $D_2O$ has a broad index dispersion, and it remains OFF-resonantly coupled to the cavity mode even beyond a mismatch of ±400 cm$^{-1}$.

Concentration-dependent experiments are done to understand the on-set of strong coupling for the OH stretching band. The coupling condition between the cavity photon and OH stretching band at different concentrations of $H_2O$ can be obtained by comparing the coupling strength ($g$) with the decay of the excited state of the molecule ($\gamma$), and the decay rate of the cavity photon ($\kappa$). When $\gamma$ and $\kappa > g$, i.e., losses in the system dominate over the energy exchange between cavity and molecule transition, termed as weak coupling regime. This condition generally does not result in the splitting of the molecular state. When $\gamma$ and $\kappa < g$, implying energy exchange between the coupled oscillators exceeds losses in the system; here, the system is said to be in strong coupling regime. Experimentally, the decay rates $\gamma$ and $\kappa$ can be directly taken as FWHM of molecular transition and cavity mode, respectively. Please note that the FWHM gives the decay rate and is inversely proportional to the overall lifetime of the state. The coupling strength ($g$) at each concentration can be calculated using experimentally obtained splitting energy ($\hbar\Omega_R$), $\gamma$ and $\kappa$ in the following relation:

$$\hbar\Omega_R = \sqrt{4g^2 - (\gamma - \kappa)^2}$$

**Table1.** Dependence of coupling strength on the concentration of $H_2O$

| Concentration (v/v%) | $\gamma$ (cm$^{-1}$) | $\hbar\Omega_R$ (cm$^{-1}$) | $g$ (cm$^{-1}$) | Coupling Regime |
|---|---|---|---|---|
| 2% | 228.4 | - | 46.7 | Weak |
| 5% | 267.2 | 150.6 | 100.2 | Intermediate |
| 7% | 276.3 | 201.4 | 123.0 | Strong |
| 9% | 294.6 | 240.1 | 144.2 | Strong |
| 11% | 310.6 | 273.5 | 161.7 | Strong |
| 15% | 338.0 | 347.6 | 201.3 | Strong |
| 100% | 463.0 | 743.1 | 406.1 | Ultra-strong |

Table 1 shows the values of $\gamma$ for different concentrations of $H_2O$ with $\kappa$ remaining ~ 135 cm$^{-1}$ (7[th] mode) throughout the experiment. Kindly note that the $\kappa$ can vary from cavity to cavity in real experiments, and we try to maintain the quality factor as closely as possible for a better comparison. Based on the values of $g$, $\gamma$ and $\kappa$, the coupling regime can be divided into three categories: weak coupling ($g^2 < \frac{(\gamma-\kappa)^2}{4}$), intermediate coupling ($g^2 > \frac{(\gamma-\kappa)^2}{4}$), and strong coupling ($2g > \frac{\gamma+\kappa}{2}$).[46] These three regimes are schematically represented in figure 5. Figure 5(a) and 5(b) shows the distinction between these three coupling regimes for different concentration of $H_2O$. The 2% $H_2O$ solution does not show any splitting of states, making it in the weak coupling regime. At 5% $H_2O$, vibro-polaritonic branches start appearing, and the system is still in the intermediate coupling regime. The system enters into strong coupling regime at all higher concentrations (above 5%). Similar experiments can't be done for $D_2O$ as the dilute solution of $D_2O$ in $H_2O$ contains a liberation band of $H_2O$ at 2100 cm$^{-1}$ which interferes in the VSC of OD stretching band of $D_2O$ at 2507 cm$^{-1}$ (figure S7).

There are few attempts to understand the nature of the vibro-polaritonic states of water molecules. The compact size of water molecules increases their density in a given concentration compared to conventional solvents.[42] Secondly, the dynamic nature of hydrogen bonding present in the system can affect spectral characteristics. The condensed phase of water has large numbers of multi-point hydrogen bonds that make the vibrational states extremely broad. Bringing a cavity mode to this envelope result in the formation of vibro-polaritonic states, but due to large refractive index dispersion, they evolve as and when the phase matching occurs. Further, the intensity of the lower polaritonic states is dampened due to the residual absorption of the broad envelope. A clear anti-crossing occurs for the 9% $H_2O$, which fits nicely with the TMM simulation. This situation changes upon increasing the concentration. This accompanies an increase in the FWHM (table 1). Kindly note that the FWHM extracted for neat water is from a non-spacer demountable cell. Using this as a reference, the TMM cavity fitting must be adjusted ~48% of the oscillator strength to match the experimental data. This indicates that ultra-strong coupling boosts the effect as expected, along with an overall increase in the self-dipolar interaction at high concentrations. Our previous findings on $CS_2$ molecules also suggest the same behavior as anticipated in the ultra-strong coupling condition.[34] Further, increasing the concentration of $H_2O$ in $D_2O$ has not affected its absorption maximum, indicating that the condensed phase of water preserves it vibrational ground state and doesn't get affected by diluting with $D_2O$ (figure S6). This is not the case if you use an aprotic solvent like DMSO, which may perturb the hydrogen bonding network.[43] Water and heavy water give very large Rabi splitting energy, and the predicted polaritonic band gap is 325 and 225 cm$^{-1}$, respectively.[47] This implies that the vibro-polaritonic states formed within this regime are more molecule-like and will be reabsorbed back into the uncoupled/dark states. Another interpretation is that the OH/OD absorption region is optically thick that won't allow the infrared photon to pass through the medium (figure S4). However, VP+ and VP- formed here is outside this energy gap (almost double), indicating that the measured Rabi splitting is a true parameter. Nevertheless, we are not sure about the comment on the very low FWHM of the VPs which may have its own origin from spectral inhomogeneity,[48] and edge absorption from the optically thick medium.[49] These issues can be addressed by looking into a 2D-IR pump probe experiment,[50] which is beyond the scope of the current work. From the static experimental observations, the vibro-polaritonic state formed in water disperses normally, and it can retain all the hybrid characteristics. Dispersion experiments and the Hopfield coefficient calculated support the above observation. Ultra-strong coupling condition invoke an appreciable variation in the self-energy term that was simply corrected by arbitrarily increasing the oscillator strength. Further, the corrected TMM fit exactly with the experimental data. A recent theoretical work study single molecule VSC and ultra-VSC of water, suggesting exactly the same behavior.[51]

In conclusion, we have systematically studied the nature of vibro-polaritonic states of water and heavy water. Concentration and pathlength-dependence studies clearly show that neat water can show all the properties of ultra-strong coupling conditions. The on-set of strong coupling is also verified by measuring the coupling strength and are categorized it into different regimes. The system enters into strong coupling above 5% of water molecules, indicating that water strong coupling can be used for testing polaritonic chemistry experiments. Such experiments reported so far show a clear envelope dependence, and vibro-polaritonic states intensity may not be the deciding factor. At the same time, inhomogeneously broad absorption characteristics of OH/OD stretching bands make it hard to test ON-resonance versus OFF-resonance effects. In our opinion, bringing a cavity mode within the FWHM is enough to achieve strong coupling, and the polaritonic branches can still possess the photon character to a large detuning condition that will benefit researchers in testing new polaritonic chemistry experiments with water.

**Acknowledgement:**

Akhila Kadyan thank IISER Mohali for research fellowship, and Jino George thanks IISER Mohali for the seed grant and the infrastructure facilities. And all this work is supported by funding from MoE, India-Scheme for Transformational and Advanced Research in Sciences (MoE-STARS/STARS-1/175).

**Figures**:

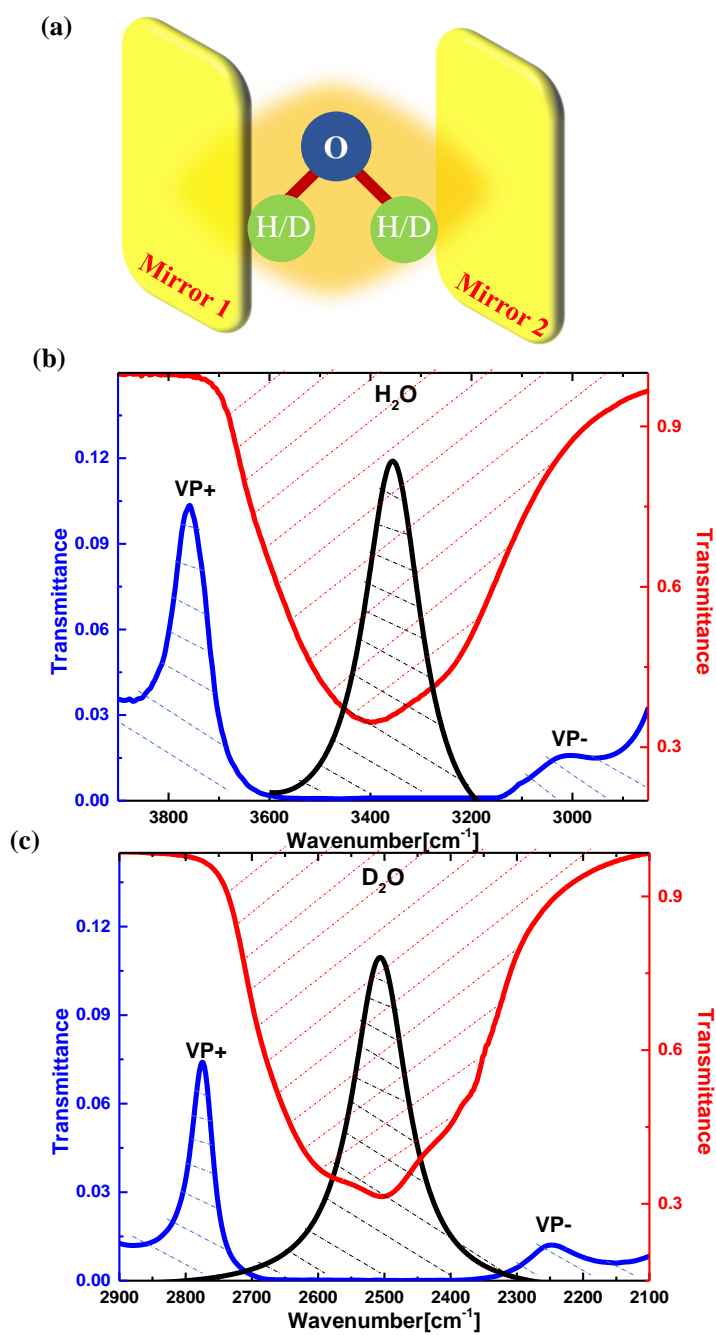

**Figure 1. (a)** Schematic representation of an FP cavity containing water/heavy water molecule. **(b)** 7$^{th}$ and **(c)** 5$^{th}$ mode of the cavity (black trace) is coupled to OH/D stretching band of $H_2O$/$D_2O$ (red trace) resulting in the formation of vibro-polaritonic states (VP+ and VP-; blue trace).

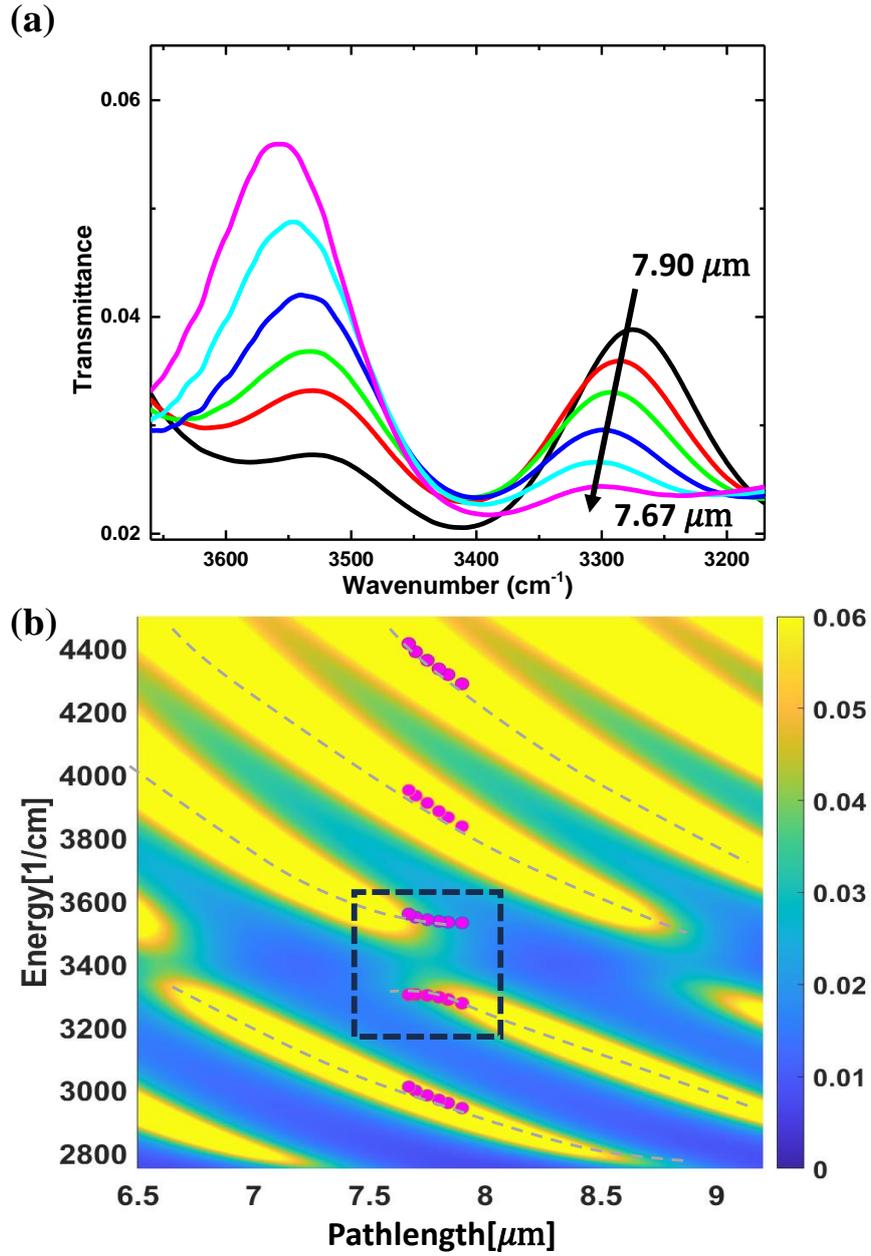

**Figure 2.** (a) Experimental dispersion of vibro-polaritonic states for 9% of $H_2O$ (in $D_2O$) coupled to $7^{th}$ mode of the FP cavity, and (b) the corresponding TMM dispersion color map with experimental points (magenta circles). The square region in the TMM is mapped experimentally as temporal evolution spectra as shown in (a).

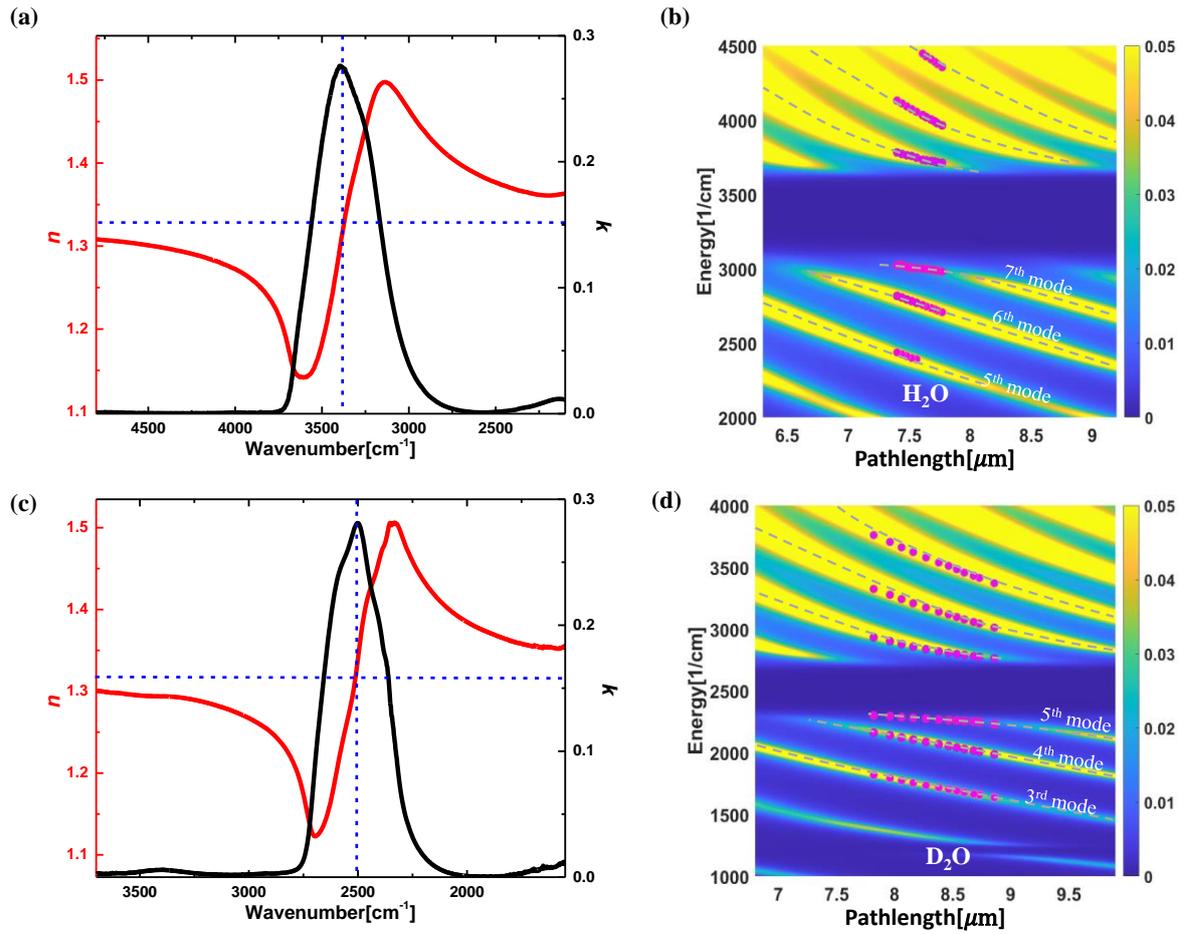

**Figure 3.** Real and imaginary parts of refractive index as a function of wavenumber for 100% pure **(a)** $H_2O$ and **(c)** $D_2O$. Dispersion of vibro-polaritonic states in 100% pure **(b)** $H_2O$ and **(d)** $D_2O$. Experimental points are incorporated as magenta circles after appropriate corrections.

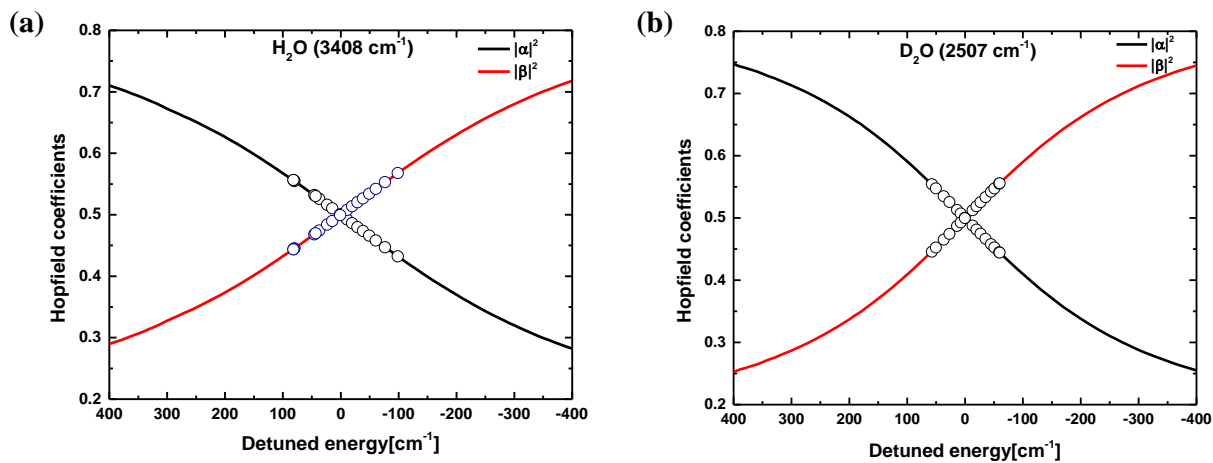

**Figure 4.** Experimentally calculated Hopfield coefficients (black hollow circles) of the lower vibro-polaritonic state for 100% pure **(a)** $H_2O$ and **(b)** $D_2O$. TMM calculation is extended to a detuning energy ($\Delta E$) of ±400 cm$^{-1}$ to indicate the trend.

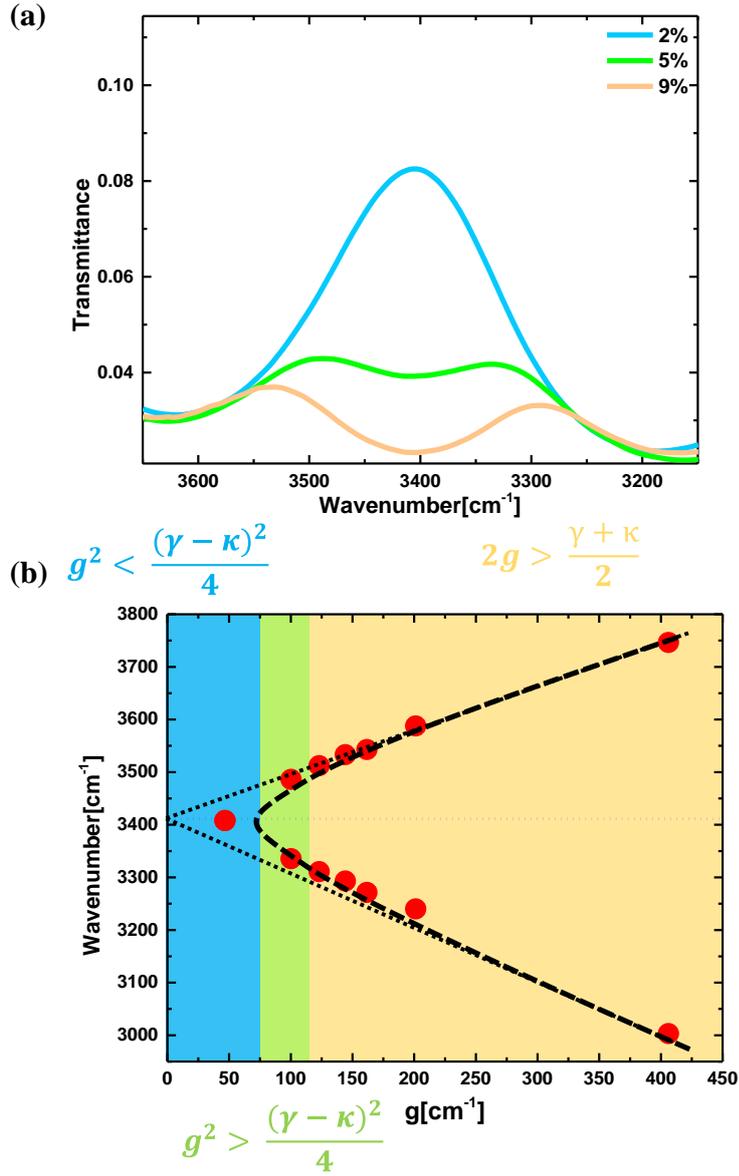

**Figure 5.** **(a)** Experimental transmission spectra for coupling of different concentrations of $H_2O$ (in $D_2O$) indicating the formation of vibro-polaritonic branches. **(b)** Evolution of vibro-polaritonic states as a function of coupling strength $g$. Different regions of light-matter coupling are represented based on numerical limits with color mapping as blue, green, and yellow corresponding to weak, intermediate and strong coupling conditions, respectively.

# Supporting Information

**Understanding the Nature of Vibro-Polaritonic States in Water and Heavy Water**

*Akhila Kadyan, Monu P. Suresh, Ben Johns and Jino George\**

*Department of Chemical Sciences, Indian Institute of Science Education and Research (IISER) Mohali, Punjab-140306, India.*
\*Email: jgeorge@iisermohali.ac.in

**CONTENTS:**

# 1. Experimental methods

*2.1 Materials and methods.*

HPLC grade $H_2O$ was purchased from Avantor chemicals and $D_2O$ was purchased from Cambridge Isotope Laboratories, Inc. $BaF_2$ windows, demountable liquid flow cell and Mylar spacers used for measurements were purchased from Specac Ltd., UK. Infrared measurements performed for both non-cavity and cavity conditions on Fourier Transform Infrared (FTIR) spectrometer (Bruker; model: INVENIO-R).

*2.2 Preparation of Fabry-Perot cavity*

A Fabry-Perot (FP) cavity is the most commonly used geometry for vibrational strong coupling experiments. It is made up of two parallelly placed metal mirrors. In this work, the FP cavity is made of 6 nm gold mirrors sputtered on two $BaF_2$ substrates and placing them parallel to one another in a demountable liquid flow cell with a 6 $\mu$m Mylar spacer.

# 2. Transfer matrix method and oscillator strength corrections

For fitting the OH/OD stretching band water in non-cavity condition with TMM simulation, the complex refractive index of water is modelled with multi-Lorentzian function[1]:

$$n(\omega) = \sqrt{n_b^2 - \sum_{j=1}^{N} \frac{f_j}{\omega^2 - \omega_j'^2 + ik\Gamma_j}}$$

where, $n_b$ is the background refractive index, $f_j$ is the strength Lorentz oscillator, $\omega_j'$ is the frequency position (in cm$^{-1}$) of Lorentz and $\Gamma_j$ is the decay rate which define the line-width of the Lorentzian. TMM fitting is done by taking 3 layers; $BaF_2$ substrate as the first layer (4mm thick; taken as semi-infinite medium) on both sides of active layer (water molecules; few micrometres thick). Figure S1 is a schematic representation of the light propagation path for a non-cavity structure used in the TMM calculation. The incident light amplitude taken as *1*, transmitted as *t* and reflected as *r*.

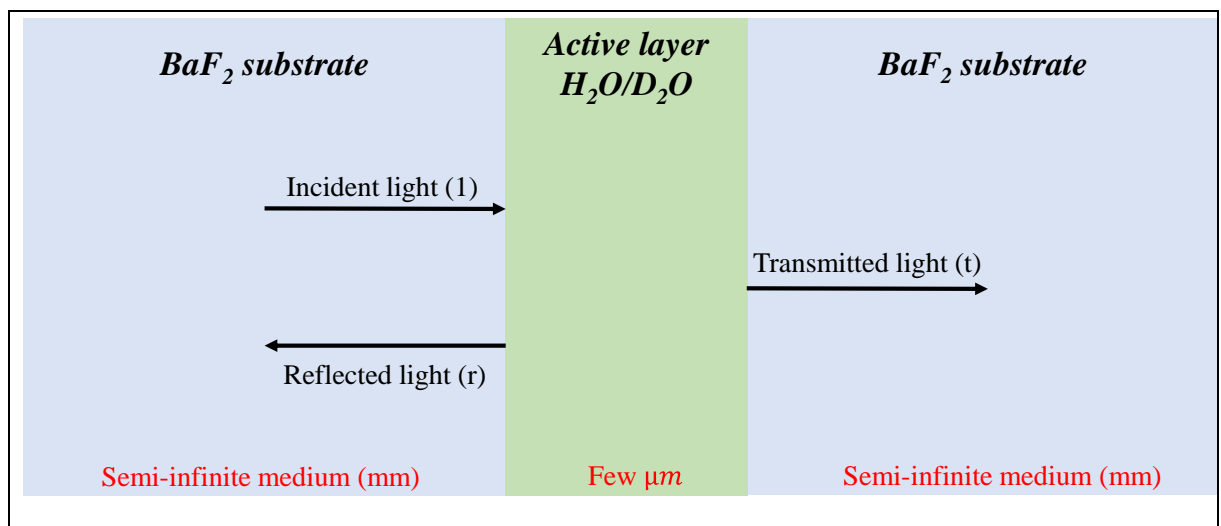

**Figure S1.** Schematic representation of the non-cavity structure used for modelling in transfer matrix methods (TMM) simulation.

The experimental non-cavity OH/OD stretching band of water obtained in a 1-2 μm thick layer of water between $BaF_2$ substrates and their corresponding multi-Lorentz fitting are shown in figure S2. Various parameters used for fitting OH/OD bands (100% water) with their respective oscillator strength, position and width are given in table S1 and table S2. After fitting the non-cavity water OH/OD stretching bands, TMM calculations are performed for the VSC of the OH/OD stretching bands inside the cavity. The cavity pathlength used in OH/OD coupling experiments is ~7.5 μm while the non-cavity OH/OD band fitting is performed with a 1-2 $\mu m$ pathlength. Further, in the TMM, the pathlength parameters are adjusted to match the experimental cavity data. Please note that the Au mirrors thickness is taken as 6 nm for all the experiments and TMM calculations; thin mirrors give a broad cavity mode ($\kappa = 135\ cm^{-1}$; for 7$^{th}$ cavity mode). When the multi-Lorentz fitting is transformed to cavity coupled situation, TMM predicted splitting energy does not match with the experimentally obtained splitting. TMM splitting energy is smaller than the experimental one. To match the TMM splitting, the oscillator strength of OH band is required to be increased which is done by increasing the strength of all the Lorentzian oscillators making OH band with a same common factor of 1.45. This correction factor is obtained by the hit and trail method. The deviation and matching of TMM without and with oscillator strength corrections, respectively are shown below (figure S3). In the case of $D_2O$, the strength of multi-Lorentzian is multiplied by a factor 1.35 to get an exact matching of TMM predicted Rabi-splitting energy with the experimentally obtained data.

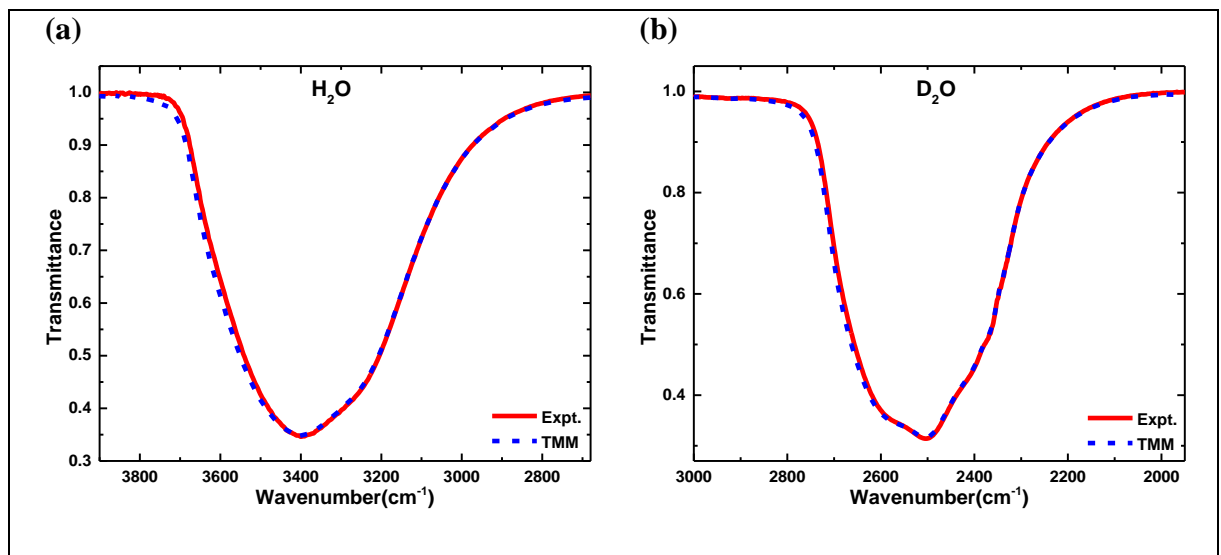

**Figure S2.** Experimental OH/OD stretching band (red curves) of **(a)** $H_2O$, and **(b)** $D_2O$ with their corresponding fitting (blue curves).

**Table S1:** Multi-Lorentz fitting parameters used for OH stretching band of $H_2O$.

| Oscillator No. | Strength ($f_j$) | Position ($\omega'_j$) | FWHM ($\Gamma_j$) |
| --- | --- | --- | --- |
| 1. | 0.544*10^3 | 2920 | 75 |
| 2. | 0.544*10^3 | 2940 | 75 |
| 3. | 1.344*10^3 | 2960 | 75 |
| 4. | 1.344*10^3 | 2980 | 75 |
| 5. | 2.144*10^3 | 3000 | 75 |
| 6. | 2.144*10^3 | 3020 | 75 |
| 7. | 7.424*10^3 | 3040 | 75 |
| 8. | 8.24*10^3 | 3080 | 75 |
| 9. | 19.664*10^3 | 3120 | 75 |
| 10. | 20.94*10^3 | 3160 | 75 |
| 11. | 3.54*10^4 | 3200 | 75 |
| 12. | 4.02*10^4 | 3240 | 75 |
| 13. | 4.4*10^4 | 3280 | 75 |
| 14. | 3.82*10^4 | 3320 | 75 |
| 15. | 1.88*10^4 | 3340 | 75 |
| 16. | 2.26*10^4 | 3360 | 75 |
| 17. | 2.3*10^4 | 3380 | 75 |
| 18. | 2.8*10^4 | 3400 | 75 |
| 19. | 1.95*10^4 | 3420 | 75 |
| 20. | 3.71*10^4 | 3440 | 75 |
| 21. | 4.1*10^4 | 3480 | 75 |
| 22. | 2.76*10^4 | 3520 | 75 |
| 23. | 2.34*10^4 | 3560 | 75 |
| 24. | 1.68*10^4 | 3600 | 75 |
| 25. | 0.01*10^4 | 3640 | 45 |
| 26. | 0.01*10^4 | 3680 | 45 |
| 27. | 0.01*10^4 | 3720 | 45 |
| 28. | 0.01*10^4 | 3760 | 45 |
| 29. | 0.01*10^4 | 3800 | 45 |

**Table S2:** Multi-Lorentz fitting parameters used for OD stretching band of $D_2O$

| Oscillator No. | Strength ($f_j$) | Position ($\omega'_j$) | FWHM ($\Gamma_j$) |
|---|---|---|---|
| 1. | 0.1*10*3 | 2140 | 55 |
| 2. | 0.1*10*3 | 2160 | 55 |
| 3. | 0.1*10*3 | 2180 | 55 |
| 4. | 0.4*10*3 | 2200 | 55 |
| 5. | 0.74*10*3 | 2220 | 55 |
| 6. | 1.04*10*3 | 2240 | 55 |
| 7. | 2.81*10*3 | 2280 | 55 |
| 8. | 9.54*10*3 | 2320 | 55 |
| 9. | 15.94*10*3 | 2360 | 55 |
| 10. | 3.8*10*4 | 2400 | 75 |
| 11. | 3.3*10*4 | 2440 | 75 |
| 12. | 3.9*10*4 | 2480 | 75 |
| 13. | 2.9*10*4 | 2500 | 75 |
| 14. | 1.2*10*4 | 2520 | 55 |
| 15. | 1.7*10*4 | 2540 | 55 |
| 16. | 1.45*10*4 | 2560 | 55 |
| 17. | 1.7*10*4 | 2580 | 55 |
| 18. | 1.59*10*4 | 2600 | 55 |
| 19. | 1.24*10*4 | 2620 | 55 |
| 20. | 1.851*10*4 | 2640 | 55 |
| 21. | 1.461*10*4 | 2680 | 55 |
| 22. | 0.21*10*4 | 2720 | 55 |
| 23. | 0.0001*10*4 | 2760 | 35 |
| 24. | 0.0001*10*4 | 2800 | 35 |
| 25. | 0.0001*10*4 | 2840 | 35 |
| 26. | 0.0001*10*4 | 2880 | 35 |
| 27. | 0.0001*10*4 | 2920 | 35 |
| 28. | 0.0001*10*4 | 2960 | 35 |
| 29. | 0.0001*10*4 | 3000 | 35 |

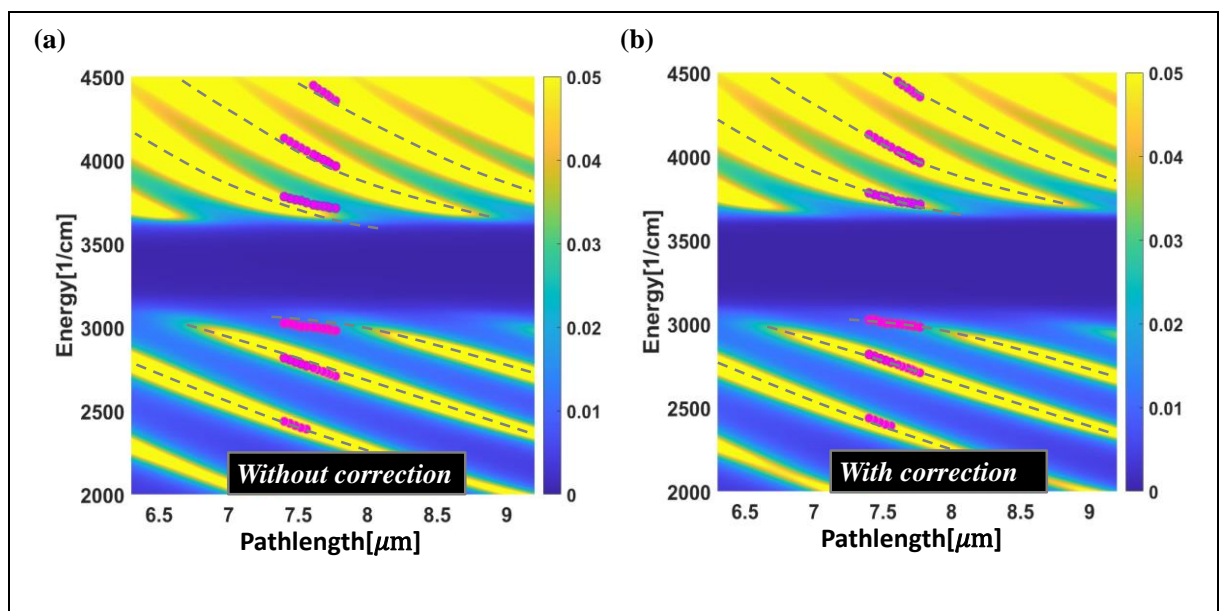

**Figure S3.** TMM dispersion of vibro-polaritonic states of $H_2O$ inside the cavity with experimental points (magenta circles) obtained by tuning the cavity mode position (7[th] mode).

## 3. Non-cavity and cavity spectra using 6 μm mylar spacer

For comparing the Rabi splitting energy with the optical saturation of bare molecular absorption, the spectra for both non-cavity and cavity conditions is obtained as shown in figure S4. In case of $H_2O$, OH band become optically thick (roughly around the absorption maximum with the transmission intensity become zero for 345 cm$^{-1}$ range) whereas the Rabi splitting energy is 740 cm$^{-1}$. This shows that the observed polaritonic bands are far away from the optical saturation length of OH stretching band. In $D_2O$, OD band saturation (optically thick) is 205

cm$^{-1}$, while Rabi splitting energy is 529 cm$^{-1}$; thus polariton bands in D$_2$O are also out of the optical band saturation region of the OD stretching band.

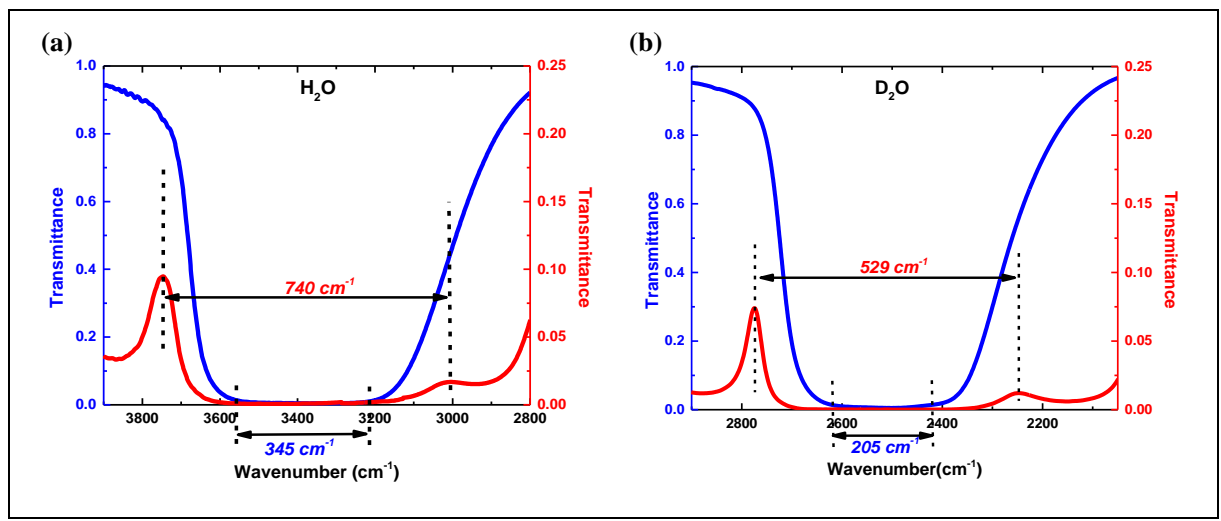

**Figure S4.** OH/OD stretching band and corresponding vibro-polariton states formation using 6 μm spacers for **(a)** H$_2$O and **(b)** D$_2$O.

## 4. Mode-dependent Rabi splitting energy

To understand the role of mode order used for VSC, 6$^{th}$, 7$^{th}$ and 8$^{th}$ order mode coupling with OH stretching band of 9% H$_2$O is performed in TMM. For different cavity pathlengths, different mode gets coupled to the OH band. For coupling OH band, cavity pathlength are estimated to be 6.5 μm, 7.68 μm, and 8.8 μm, and are further used in the TMM calculations. Please note that 7$^{th}$ cavity mode is only used for experimental studies due to technical difficulties. Therefore, TMM calculations are used for getting information on mode dependence on Rabi splitting energy. As shown in figure S5 and table S3, the Rabi splitting energy does not change with the order of the cavity mode used for VSC.

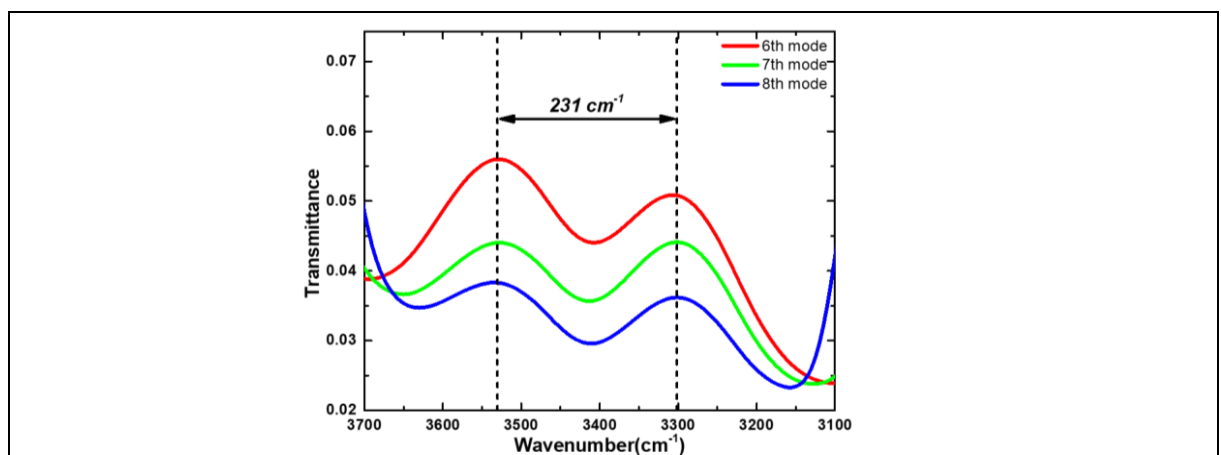

**Figure S5.** Cavity mode dependence on Rabi splitting energy calculated from TMM for VSC of OH band of 9% H$_2$O diluted in D$_2$O.

**Table S3:** Output parameters obtained from TMM calculation for coupling 9% $H_2O$ diluted in $D_2O$.

| Cavity pathlength (μm) | Coupled mode order | Mode Q-factor | Rabi-Splitting (cm$^{-1}$) |
|---|---|---|---|
| 6.53 | 6$^{th}$ | 24 | 231 |
| 7.68 | 7$^{th}$ | 29 | 231 |
| 8.80 | 8$^{th}$ | 32 | 231 |

## 5. Concentration-dependent studies for $H_2O$

For performing concentration-dependent experiments with $H_2O$, the dilution of $H_2O$ is done with $D_2O$. For both non-cavity and cavity experiments, a 6 μm Mylar spacer is used between the $BaF_2$ substrates. In non-cavity infrared spectra, the OH stretching band does not show any shift in the position indicating the H-bonding network in $H_2O$ in not getting affected by its dilution with $D_2O$ (figure S6(a)). In cavity experiments, 2% $H_2O$ solution does not show any splitting making the system in the weak coupling regime. For 5% solution, band splitting starts to take place but is not sufficient enough to achieve strong coupling. This regime is taken as intermediate coupling regime which is just the on-set of strong coupling. All concentration at and above 7% of water clearly follow the strong coupling conditions (figure S4(b)). This indicated that even a 7% $H_2O$ solution can be used for performing strong coupling experiments and making $H_2O$ one of the most useful contenders to study the effect of VSC on many physical and chemical processes.

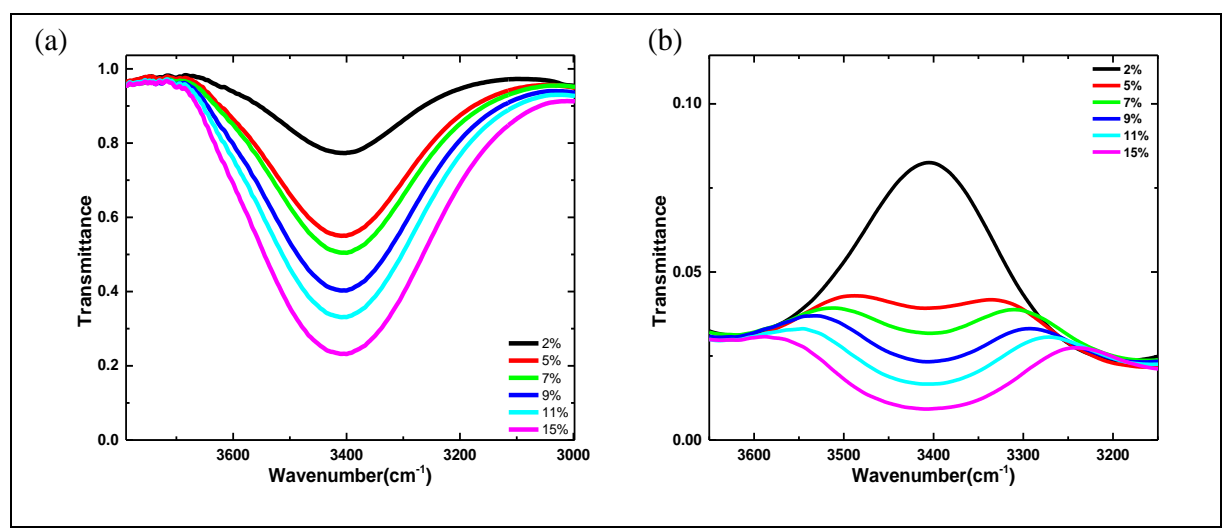

**Figure S6.** Concentration-dependent infrared spectra of $H_2O$ in **(a)** non-cavity and **(b)** cavity conditions.

## 6. Concentration-dependent studies for $D_2O$

Similar to $H_2O$, concentration-dependent studies are tried with $D_2O$ by diluting it with $H_2O$. In the case of non-cavity, the OD stretching band doesn't show any frequency shift with dilution but it contains a $H_2O$ liberation band nearby (figure S7). $H_2O$ liberation band interferes with the cavity OD stretching band VSC experiments. Here, multi-mode coupling occurs in which the VP- of the OD stretch mix with the liberation band of $H_2O$. As a result, coupling strength dependence of the OD stretching band cannot be done with $H_2O$ dilution.

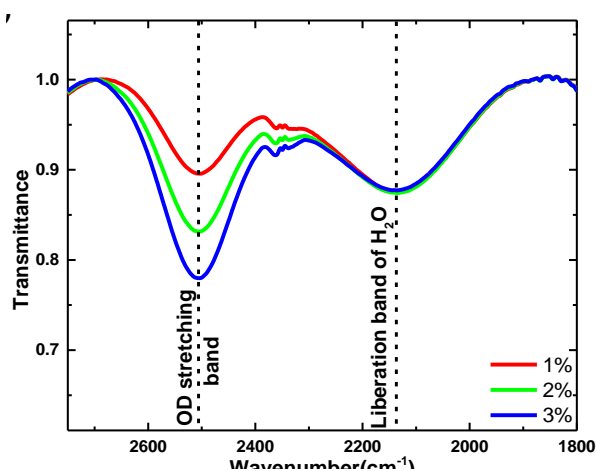

**Figure S7.** Concentration-dependent infrared spectra of $D_2O$ in $H_2O$ show the $H_2O$ liberation band is close by the OD stretching band.